\documentclass[amssymb,useAMS,prd,aps,amsmath,twocolumn,superscriptaddress,nofootinbib]{revtex4}
\usepackage{pslatex}
\usepackage{graphicx}
\usepackage{psfrag}

\usepackage[usenames]{color}
\usepackage{color,graphicx,epsfig,amsmath,amssymb}

\newcommand{\beq}{\begin{equation}}
\newcommand{\eeq}{\end{equation}}
\newcommand{\ben}{\begin{eqnarray}}
\newcommand{\een}{\end{eqnarray}}
\newcommand{\bi}{\begin{itemize}}
\newcommand{\ei}{\end{itemize}}
\newcommand{\nn}{\nonumber}

\newcommand{\ghost}[1]{ }

\newcommand{\myslash}[1]{\slash\!\!\!\!{#1}}

\begin{document}

\title{Clarifying the covariant formalism for the SZ effect due to 
  relativistic non-thermal electrons}

\author{C\'eline B\oe hm}
\affiliation{LAPTH, UMR 5108, 9 chemin de Bellevue - BP 110,
  74941 Annecy-Le-Vieux -- France}
\email{celine.boehm@lapp.in2p3.fr}

\author{Julien Lavalle}
\affiliation{Dipartimento di Fisica Teorica,
  Universit\`a di Torino - INFN,
  via Giuria 1,
  10125 Torino -- Italia}
\email{lavalle@to.infn.it}

\date{today}

\begin{abstract}
  We derive the covariant formalism associated with the relativistic
  Sunyaev-Zel'dovich effect due to a non-thermal population of high energy
  electrons in clusters of galaxies. More precisely, we show that the
  formalism proposed by Wright in 1979, based on an empirical approach
  to compute the inverse Compton scattering of a population of relativistic 
  electrons on CMB photons, can actually be re-interpreted as a Boltzmann-like 
  equation, in the single scattering approximation.
\end{abstract}


\maketitle

\begin{flushleft}
Preprint: DFTT-22/2008
\end{flushleft}

\section{Introduction}
\label{sec:intro}

The purpose of this paper is (i) to revisit the formulation of the relativistic
Sunyaev-Zel'dovich (rSZ) distortion of the cosmic microwave background (CMB)
due to a population of non-thermal high energy electrons in clusters of 
galaxies, and (ii) to provide a unified and more comprehensive framework to 
compute the rSZ effect.

Inverse Compton scattering of non-relativistic thermal electrons on cosmic
microwave background (CMB) photons has been widely investigated in the
literature since the 1960-1970's, notably in the context of the seminal
works by Sunyaev and Zel'dovich on the SZ
effect~\cite{1972CoASP...4..173S,1980ARA&A..18..537S}, and in early
studies on relativistic electron cosmic rays in high energy astrophysics
(e.g.~\cite{1965PhRv..137.1306J,1968PhRv..167.1159J,1970RvMP...42..237B}). The
SZ effect generated by a population of  hot (still non-relativistic) electrons
has also been addressed in detail since the 1980's (see
e.g.~\cite{1979ApJ...232..348W,1989ApJ...339..619T,1995ApJ...445...33R,
1995ARA&A..33..541R,1998ApJ...499....1C,1998ApJ...502....7I,1998ApJ...508....1S,
1997astro.ph..9065S,2001ApJ...554...74D}), while the SZ 
effect originating from relativistic electron cosmic rays has been discussed 
in e.g. Refs.~\cite{1999ApJ...523...78M,1999PhR...310...97B,2000A&A...360..417E,
2003A&A...397...27C}. So far, these works on hot thermal or relativistic
non-thermal electrons have used different approaches.

For example, in  Refs.~\cite{1998ApJ...499....1C,1998ApJ...502....7I,
1998ApJ...508....1S,1997astro.ph..9065S,2001ApJ...554...74D}, the authors have 
computed the relativistic corrections to the standard thermal SZ predictions 
due to the hot trail of intracluster thermal electrons (including the effect 
from multiple scattering). They used a Fokker-Planck expansion of the 
collisional Boltzmann equation in terms of the electron temperature-to-mass 
ratio $T/m\lesssim 0.1$, in which the full squared matrix amplitude of Compton 
scattering (that is given in any textbook, e.g.~\cite{Peskin:1995ev}) was 
implemented for the relevant kinematics. Such an approach therefore 
not only encompasses but also extends the usual calculation of the standard 
thermal SZ effect, thus making this formalism very attractive. Nonetheless, it 
has not been applied yet to the case where there are two populations of 
electrons (one dominant semi-relativistic population and a subdominant 
relativistic population).

The authors of e.g. 
Refs.~\cite{1979ApJ...232..348W,1989ApJ...339..619T,1995ApJ...445...33R,
1999ApJ...523...78M,2000A&A...360..417E,2003A&A...397...27C}
have considered the SZ contributions coming from a population of thermal 
electrons plus an extra source of non-thermal electron cosmic rays, as we want 
to do. However, they used a more empirical approach based on a classical 
radiative transfer method proposed by 
Chandrasekhar~\cite{1950ratr.book.....C}. Indeed, they computed the spectral 
shift in the CMB photon intensity due to Compton scattering processes with 
electrons, scattering \emph{off} some photon frequencies (from energy $E$ to 
$E'$) and scattering \emph{in} some others (from energy $E'$ to $E$). The 
squared matrix amplitude (that is the probability of frequency change) was 
expressed in the electron rest frame, leading to an expression close to its 
associated non-relativistic limit, such as the one originally used by 
Chandrasekhar. Multiple scatterings were eventually implemented by further 
using Chandrasekhar's radiative approach.

In Sect.~\ref{sec:boltz}, we demonstrate that Chandrasekhar's radiative 
transfer method and the Boltzmann-like approach mentioned above are in fact 
formally equivalent in the single scattering approximation. They thus give the 
same results, which we show by using the full relativistic squared matrix 
amplitude instead of the photon frequency redistribution probability function 
derived by Wright~\cite{1979ApJ...232..348W}. This therefore sketches a unified 
formalism for the computation of both hot thermal and relativistic non-thermal 
SZ effects, as subsequently found in~\cite{2009arXiv0902.2595N}. For 
illustration, we derive the relativistic photon energy redististribution 
function in the CMB reference frame (therefore different from Wright's 
$P(s,\beta)$ function) that can be used to compute the relativistic SZ 
effect generated by an additional population of high energy electron cosmic 
rays in a cluster. For the sake of completenes, we revisit in 
Sect.~\ref{sec:compton} the full relativistic phase space and squared 
amplitude of the inverse Compton process in the relativistic limit. Our 
conclusions are drawn in Sect.~\ref{sec:concl}. Numerical applications of this 
framework will be given in a companion paper~\cite{barthes_etal_in_prep}.

\section{Equivalence between radiative transfer and the Boltzmann equation}
\label{sec:boltz}

The purpose of this section is to demonstrate the equivalence between the 
approach based on the Boltzmann equation and Chandrasekhar's radiative 
transfer method to compute the rSZ effect, in the single scattering 
approximation.

\subsection{Scattered intensities in the radiative transfer approach}
\label{subsec:transfer}
Chandrasekhar's radiative transfer method has been widely used in the
literature, notably to compute the SZ effect due to non-thermal electrons
in clusters of galaxies (for a review, see~\cite{1999PhR...310...97B}). The 
principle of this approach is recalled in the appendix 
(see Sect.~\ref{app:wright_formalism}), as well as the precise derivation of 
the CMB intensity deviation in the relativistic regime and in the single 
scattering approximation (see Sect.~\ref{app:in_out}). Here we 
summarize the main results coming out of this approach.

The spectral shift in the CMB intensity due to scatterings with relativistic 
electrons can be expressed as the difference between two contributions. The 
first one is associated with the photons scattered \emph{out} 
($\widehat{I}_\gamma^{\rm s-out}(E_k)$) from the CMB spectrum, from energy $E_k$ 
to energy $E_k'\neq E_k$, and the second is associated with the photons 
scattered \emph{in} ($\widehat{I}_\gamma^{\rm s-in}(E_k)$) from energy $E_k'$ 
to energy $E_k$ (after averaging over the direction of the incoming photons): 
\ben
\Delta I_\gamma(E_k) = \widehat{I}_\gamma^{\rm s-in}(E_k) -
\widehat{I}_\gamma^{\rm s-out}(E_k)\;,
\een
The expressions of the \emph{in} and \emph{out} intensities can then be 
calculated. They are given in Sect.~\ref{app:in_out} (see 
Eqs.~\ref{eq:app_out} and \ref{eq:app_in}) as integrals over the line of 
sight $dl$:
\ben
\widehat{I}^{\rm s-out}(E_k) =  \int dl
\int \frac{d^3\vec{p}} {(2 \pi)^3} f_e(E_p) \int \frac{d\Omega_{k'}}{32 \pi^2}
\frac{ t^2 \ |{\cal M}|^2 }{(1-\beta \mu)}  \frac{I_\gamma^0(E_k)}{E_p^2} \nn
\een
and
\ben
\widehat{I}^{\rm{s-in}} (E_{k}) = \int dl \int \frac{d^3 p}{(2 \pi)^3}
f_e(E_p) \int \frac{d\Omega_{k'}}{32\pi^2} \frac{|\widetilde{\cal M}|^2}{E_p^2}
\frac{I_\gamma^0(\tilde{t} E_k)}{\tilde{t}(1-\beta\mu)} \nonumber
\een
 where $I_\gamma^0$ is the intensity associated with the photon black-body 
spectrum (before interaction with the electrons) and $f_e$ is the energy 
distribution function characterizing the relativistic electrons (of energy 
$E_p$). The term $|{\cal M}|^2$ is the squared matrix amplitude associated with
the process $(p,k\rightarrow p',k')$ for the scattered-out part, and
$|\widetilde{\cal M}|^2$ describes the process $(p,k'\rightarrow p',k)$ for 
the scattered in part. The terms $t$ and $\tilde{t}$ feature the photon 
energy transfer for both processes, respectively, and are defined in 
Eqs.~(\ref{eq:ekprime},\ref{eq:eqtt}). $\mu$ characterizes the angle between 
the incoming electron of momentum $\vec{p}$ and the photon of momentum 
$\vec{k}$ (see Eq.~\ref{eq:def_angles}).

As we will show in the next section, the photon frequency redistribution 
probability function considered by Wright in 1979~\cite{1979ApJ...232..348W}, 
which has been continuously used afterwards for rSZ predictions 
\cite{1995ApJ...445...33R,1999ApJ...523...78M,2000A&A...360..417E,
2003A&A...397...27C}, will differ from our analysis in that it was derived in 
the electron rest frame, but both are consistent~~\cite{2009arXiv0902.2595N}. 
Anyway, the equivalence between the radiative transfer and covariant 
approaches --- in the relativistic regime and single 
scattering approximation --- can be formally demonstrated without the need 
for replacing the squared matrix amplitude by its expression.

\subsection{Boltzmann-like formalism} 

\subsubsection{Covariant collisional Boltzmann equation}
\label{subsec:boltz}

The theoretical framework developed in~\cite{1998ApJ...499....1C,
1998ApJ...502....7I,1997astro.ph..9065S,2001ApJ...554...74D} to treat the
semi-relativistic corrections of the thermal SZ is the covariant
collisional Boltzmann equation (see e.g.~\cite{1970mtnu.book.....C,
1990eaun.book.....K,2002rbet.book.....C,2003_escobedo}), which
expresses the momentum transfer between different interacting species.

When focusing on photons scattered by a gas of electrons in a covariant
scheme, the spatial convective currents are irrelevant, and the time
evolution of the photon phase space is given by the covariant collisional
Boltzmann equation as:
\ben
\frac{d}{dt} f_\gamma(E_k) &=& {\cal I}_{\rm coll}(E_k)\;.
\label{eq:boltz_def}
\een
The collision integral is defined by:
\ben
{\cal I}_{\rm coll}(E_k) &\equiv& \frac{1}{2 E_k} \int \prod_{X = p,p',k'}
\frac{d^3\vec{X}}{(2\pi)^3 2 E_X} \; \delta^4(p+k-p'-k')\;
\times \nn\\
&\times & |{\cal M}|^2 \, \bigg\{ f_\gamma(E_k)f_e(E_p) - f_\gamma(E_{k'})
f_e(E_{p'}) \bigg\}\nn\\
&=& - \int \frac{d^3\vec{p}}{(2\pi)^3}
\; d{\rm LIPS} \; \frac{|{\cal M}|^2}{4 E_k E_p} \times \nn\\
&\times & \bigg\{ f_\gamma(E_k)f_e(E_p) - f_\gamma(E_{k'})f_e(E_{p'}) \bigg\}\;,
\label{eq:boltz1}
\een
where $f_\gamma$ and $f_e$ are the photon and electron distributions,
respectively, and where we have neglected the Pauli blocking and stimulated
emission factors (even if the latter were included, they would automatically 
disappear as soon as treating relativistic electrons, since in this case 
$f_e(E_p)\simeq f_e(E_{p'})$, as we will show in the following). This equation 
features the Lorentz invariant phase space $d{\rm LIPS}$ and the squared 
matrix amplitude $|{\cal M}|^2$ associated with the process 
$(p,k\leftrightarrow p',k')$. These will be made explicit in 
Sect.~\ref{sec:compton}. Note that this formalism rests on the 
microreversibility of the process.

Let us now discuss the meaning of the electron distribution featured
above. As we have already stressed, we are considering two different
populations of electrons, one thermal and the other non-thermal. The former is
assumed dominant, but we are interested in the subdominant population only.
Formally, the above electron distribution $f_e$ characterizes the sum of these 
two populations, that is $f_e = f_e^{\rm n-th} + f_e^{\rm th}$. Hence, the 
question of whether we can separate their respective contributions arises. 
When integrating over the electron momenta, we see that at low energy, say 
below 1 MeV typically, only the thermal part $f_e^{\rm th}$ will be relevant 
in the calculation, while at higher energy, only the non-thermal part will 
contribute. Since we restrict ourselves to the single scattering limit, we can 
therefore safely separate those two contributions since none of them will 
affect the other one. Consequently, in the following, we will only deal with 
the relativistic non-thermal part, and will set $f_e = f_e^{\rm n-th}$ for 
convenience. Note that in the multiple scattering case, this would not hold 
anymore, since there would be a mixing of the SZ effects coming from both 
populations of electrons: for instance, a photon could first be scattered by a 
thermal electron, and then by a non-thermal one. On that account, we would 
have to treat both populations in a self-consistent manner. Nevertheless, such 
a complex analysis is beyond the scope of this paper, and probably too 
sophisticated regarding the small contribution to the SZ effect that is 
expected from relativistic electrons.

The collisional Boltzmann equation written above describes the time
evolution of the photon phase space distribution due to elastic collisions
with electrons. The electron distribution has therefore to be known
at any time. In the following (Sect.~\ref{subsec:rsz}), we will demonstrate
that if one assumes that the electron phase space density does not change
significantly over one (or even several) characteristic travel time of a
photon through a cluster, then one can use this formalism to treat the SZ
effect originating from non-thermal electrons in the single scattering
approximation. The above hypothesis is justified since, if we take a typical 
scale of 10 Mpc for a cluster, then the mean time for a photon to cross it is 
$\sim 30$ Myr while the typical timescale for the relativistic electron energy 
loss by inverse Compton scattering on CMB photons is $\sim 300$ Myr.

\subsubsection{Covariant formalism applied to the rSZ effect}
\label{subsec:rsz}
Let us consider a cluster containing a tiny population of electrons, and track 
the photon distribution. The integration over the time, 
in Eq.~(\ref{eq:boltz_def}), translates to an integration over a geodesic 
($dl = c\ dt$):
\ben
\int d f_\gamma(E_k) &=& - \int dl \int \frac{d^3\vec{p}}{(2\pi)^3 }
\; d{\rm LIPS} \; \frac{|{\cal M}|^2 }{4 E_k E_p} \times \nn\\
&\times & \bigg\{ f_\gamma(E_k)f_e(E_p) - f_\gamma(E_{k'})f_e(E_{p'}) \bigg\}\;.
\label{eq:df_meaning}
\een
If we consider the electron density to be small enough, we can
assume that a CMB photon has a very small probability to experience Compton
scattering with electrons along its path through a cluster. In this case, the
single scattering approximation is fully relevant and gives:
$\int d f_\gamma = f_\gamma^{\rm after}-f_\gamma^{\rm before}$, where
$f_\gamma^{\rm before}$ is the original black-body spectrum distribution
before the CMB photons enter into the cluster, namely $f_\gamma^0$, and
$f_\gamma^{\rm after}$ is the photon distribution after travelling through the 
cluster and experiencing one scattering. An observer looking in the direction 
of the cluster will therefore observe $f_\gamma^{\rm after}=f_\gamma\neq 
f_\gamma^0$.
Expressing now this equation in terms of the photon
intensity, that is $I_\gamma(E) = E^3 f_\gamma(E)$, we obtain:
\ben
I_\gamma (E_k) &=& I_\gamma^0 (E_k) - \int dl \int \frac{d^3\vec{p}}{(2\pi)^3 }
\; d{\rm LIPS} \; \frac{|{\cal M}|^2}{4 E_k \ E_p}  \times\nn\\
&\times &  \left\{ I_\gamma(E_k)f_e(E_p,\vec{x}) -
\frac{E_k^3}{E_{k'}^3}\ I_\gamma(E_{k'})f_e(E_{p'},\vec{x}) \right\}\;,
\label{eq:boltz2}
\een
where $I_\gamma^0$ is the intensity of the pure black-body spectrum of the
CMB before entering in the cluster. The first term in the sum under
the integral characterizes the frequencies shifted out from $E_k$ to $E_{k'}$,
while the second term characterizes the frequencies shifted in from $E_{k'}$
to $E_k$, both due to Compton scattering. Note that the electron distribution
$f_e$ is not ubiquitous anymore, and must now also carry information of the
spatial distribution, which will be relevant for the line of sight integral
along $dl$.

As we stated previously, if the number density of the relativistic electron
population is very small, then the interactions between the CMB
photons and this population is well approximated by the single scattering
approximation. This hypothesis further enables us to treat several populations
of electrons (e.g. a dominant, thermal and/or semi-relativistic, population
and a subdominant relativistic population) separately.

Let us now specifically focus on the relativistic population. As we are 
considering electrons with typical energies $E_p\gtrsim 1$ MeV and CMB photons 
with energies $E_k \lesssim 10^{-3}$ eV, we can safely make use of the limit
$\alpha\equiv E_k/E_p \rightarrow 0$ in the above equation, which is fully 
valid in the whole range of the CMB photon energy distribution in the single 
scattering approximation. Within this limit, due to energy-momentum 
conservation in the Compton scattering, we will show in 
Sect.~\ref{sec:compton} that $E_{p'}\rightarrow E_p$, such that we can replace 
$f_e(E_{p'})$ by $f_e(E_p)$ (see Eqs.~\ref{eq:epprime} and~\ref{eq:dlips3}). 
This is the key point to understand why the Boltzmann framework can be linked 
to the radiative transfer approach for the SZ effect associated with non-thermal
electrons. Most importantly, in the single scattering approximation, the photon
intensity $I_\gamma$ under the integral of the right hand side can be replaced
by that corresponding to the pure black-body spectrum $I_\gamma^0$. Defining
$\Delta I_\gamma \equiv I_\gamma - I_\gamma^0$, we obtain:
\ben
\Delta I_\gamma (E_k) &=& -2\int dl \int \frac{d^3\vec{p}}{(2\pi)^3 }
f_e(E_p,\vec{x}) \; d{\rm LIPS}  \nn\\
&\times &  \frac{|{\cal M}|^2}{4 E_k \ E_p} \;
\left\{ I_\gamma^0 (E_k) - \frac{\ I_{\gamma}^0 (E_{k'})}{t^3}\right\}\;,
\label{eq:boltz3}
\een
where $t\equiv E_{k'}/E_k$, and where we have explicitly added a factor of two
to account for the sum over the two photon polarizations (so that it should
not appear in the definition of $I_\gamma^0$ in the right hand side). This
expression fully characterizes the intensity after removing the contribution
of photons scattered out (from $E_k$ to $E_{k'}$) and after adding that of
photons scattered in (from $E_{k'}$ to $E_k$), exactly like \`a la Wright. The 
term proportional to $I_\gamma^0(E_k)$ is the
scattered \emph{out} contribution, and the term proportional to 
$I_\gamma^0(E_{k'})$ is the scattered \emph{in} contribution. Hence 
Chandrasekhar's formalism, when applied to the relativistic SZ effect, as 
recalled in the previous section, is 
strictly equivalent to using the Boltzmann equation (in the single scattering 
approximation) --- detailed expressions related to Chandrasekhar's formalism 
are derived in the appendix, see Sect.~\ref{app:in_out}. We can rewrite the 
above equation in terms of the differential Compton cross section, after 
performing the integrals over $d^3\vec{p'}$ and $dE_{k'}$:
\ben
\label{eq:boltz4}
\Delta I_\gamma (E_k) &=& -2 \int dl \int \ \frac{d^3\vec{p}}{(2\pi)^3}
f_e(E_p,\vec{x})  \\
&\times& \int d\Omega_{k'} \; \beta_{\rm rel} \;
\frac{d \sigma}{d \Omega_{k'}}
\left\{I_{\gamma}^0 (E_k) -\frac{I_{\gamma}^0(tE_{k})}{t^3}\right\}\;,\nn
\een
where $\beta_{\rm rel}$ is the relative velocity between the
incoming particles (see Eq.~\ref{eq:def_sigma}), where $d \Omega_{k'}$ is the
solid angle associated with photons of energy $E_{k'}\neq E_k$, and
where the differential Compton cross section $d\sigma$ is defined in
Eq.~(\ref{eq:def_sigma2}). We will sketch the full angular dependence in the
next subsection.

\subsection{Equivalence between the radiative transfer and covariant approaches}
\label{subsec:equi}

Eqs.~(\ref{eq:boltz3}) or (\ref{eq:boltz4}) contain two contributions, which
can conveniently be expressed as:
\ben
\Delta I_\gamma (E_k) =   I_\gamma^{\rm in}(E_k) - I_\gamma^{\rm out}(E_k),
\label{eq:I_in_I_out}
\een
where the upper script \emph{out} characterizes the part of the spectrum which
is shifted out from energy $E_k$ to other energies, while \emph{in} stands for
other frequencies which are shifted in to energy $E_k$ because of scattering.
According to Eq.~(\ref{eq:boltz4}), the term $I_\gamma^{\rm out}(E_k)$ is given
by:
\ben
I_\gamma^{\rm out}(E_k) &=& 2 \int dl \int \ \frac{d^3\vec{p}}{(2\pi)^3}
f_e(E_p,\vec{x})  \nn\\
&\times& \int d\Omega_{k'}\; \beta_{\rm rel} \; \frac{d \sigma}{d \Omega_{k'}} \;
I_\gamma^0 (E_k)\;,
\label{eq:boltz_out}
\een
where we recall that the factor of two in front of the right hand side
accounts for the sum over the two photon polarizations (see
Eq.~\ref{eq:boltz3}). This can be rewritten as:
\ben
I_\gamma^{\rm out}(E_k) = 2 \; \tau \;  I_\gamma^0(E_k) \;,
\label{eq:I_out_tau}
\een
where we define:
\ben
\tau &\equiv& \int dl \int \ \frac{d^3\vec{p}}{(2\pi)^3} f_e(E_p,\vec{x})
\int d\Omega_{k'} \; \beta_{\rm rel} \; \frac{d \sigma}{d \Omega_{k'}}\\
&=& \int dl \int \ \frac{d^3\vec{p}}{(2\pi)^3} f_e(E_p,\vec{x})
\int \frac{d\Omega_{k'}}{64\pi^2} \frac{t^2|{\cal M}|^2}{E_p^2 (1-\beta\mu)}\;.
\nn
\label{eq:def_tau}
\een
We have used the definition of the differential cross section from
Eq.~(\ref{eq:sigma_angle}), and that of the relative velocity
$\beta_{\rm rel}=(1-\beta\mu)$. Now we need to average over the solid angle
$d\Omega_k= d\phi d\mu$ associated with the incoming photons, so that the
actual measured shifted intensity will be:
\ben
\widehat{I}_\gamma^{\rm out}(E_k) &=& 2 \; \widehat{\tau} \; I_\gamma^0(E_k)\\
\widehat{\tau} &\equiv& \int\frac{d\Omega_k}{4\pi} \tau\nn\;.
\een
If we perform the integral over $d\Omega_{k'} = d\phi' d\mu'$ appearing 
in the definition of $\widehat{\tau}$, we obtain the full relativistic 
(inverse) Compton cross section given in~Eq.~(\ref{eq:sigma_int}),
which still depends on $\mu$, $E_k$ and $E_p$. Then, we can perform the
average over the ingoing photon solid angle $d\Omega_k = 2\pi d\mu$.
The result is found to depend on powers of $\alpha\gamma^2 = \gamma E_k /m$.
Though for CMB photons $\alpha\gamma^2$ is negligible up to ultra-relativistic
electron energies, that is as long as $\gamma\lesssim 10^8$, we still give the
result at next-to-leading order (see Eq.~\ref{eq:lim_sigma_int}):
\ben
\widehat{\tau} &=& \int dl \int   \frac{d^3\vec{p}}{(2\pi)^3} f_e(E_p,\vec{x})
\sigma_{\rm T}
\left(  1 - \frac{8}{3}\alpha\gamma^2 +\frac{2\alpha}{3}  \right) \;,
\een
where $\sigma_{\rm T}$ is the Thomson scattering cross section. Note that it is
quite consistent to find a result in which the cross section is close
to the Thomson cross section for the diffused \emph{out} term, since we are
in the limit $\gamma E_k\ll m$. Indeed, this corresponds
exactly to seeing a low energy photon in the electron energy rest frame: in
this sense, as very well known, we are in the Thomson regime, while not
dealing with a Thomson diffusion at all. If we had found other dependences on
angles under the integral, it would have been more difficult to integrate the 
cross section out in the CMB rest frame, as it will be shown to be the case 
for the scattered \emph{in} a bit further. Note finally that a mere 
transformation of the above expression by re-expressing the zenith angles 
of the photons $\mu$ and $\mu'$ in the electron rest frame would have resulted 
in a more convenient form, as it is well explained 
in~\cite{2009arXiv0902.2595N}.

We recall that the spatial dependence of the electron population has to be
included in the distribution $f_e$. $\tau$ can be, as displayed above,
interpreted as the optical depth associated with the electron population. To
provide an expression closer to the standard lores, we can assume that the
energy and the spatial distributions of electrons can be
factorized, such that $f_e(E_p,\vec{x}) = n_e(\vec{x}) \tilde{f}_e(E_p)$,
where $\tilde{f}_e$ is the normalized momentum distribution
($\int d^3\vec{p}/(2\pi)^3 \tilde{f}_e(E_p) = 1$). This naturally gives
$\int d^3\vec{p} /(2\pi)^3 f_e(E_p,\vec{x}) = n_e(\vec{x})$, to which one can
stick instead of the previous trick, depending on one's convenience. We can
therefore rewrite the previous equation as:
\ben
\widehat{\tau} & = & {\cal K} \; \tau_{\rm nr}
\label{eq:tau_nr}
\een
where:
\ben
{\cal K}  &\equiv & \int \frac{d^3\vec{p}}{(2\pi)^3}
\tilde{f}_e(E_p)
\left(  1 - \frac{8}{3}\alpha\gamma^2 -\frac{2\alpha}{3}  \right) \;,\\
{\rm and}\;\;    \tau_{\rm nr} &\equiv& \int dl \; \sigma_T \;
n_e(\vec{x})\;.\nn
\label{eq:corr_tau_nr}
\een
We thus recover the standard definition of the optical depth
$\tau_{\rm nr}$ for non-relativistic thermal electrons when neglecting the
terms proportional to $\alpha$ and $\alpha\gamma^2$. The factor
${\cal K}\lesssim 1$ has to be considered as a relativistic correction to
apply to $\tau_{\rm nr}$. Finally, we remind that this optical depth will have
to be averaged over the spatial resolution of the telescope for quantitative
prediction purposes (see Eq.~\ref{eq:res_ave}).

The scattered \emph{in} term of Eq.~(\ref{eq:I_in_I_out}),
$I_\gamma^{\rm in}(E_k) $, is given, according to Eq.~(\ref{eq:boltz4}), by:
\ben
I_\gamma^{\rm in}(E_k) &=&  2 \int dl \int \frac{d^3\vec{p}}{(2\pi)^3}\
f_e(E_p,\vec{x})\times\nn \\
&\times & \int d\Omega_{k'} \;  \beta_{\rm rel} \;
\frac{d \sigma }{d \Omega_{k'}}  \; \frac{I_{\gamma}^0 (t E_{k})}{t^3}\nn\\
&=& 2 \int dl \int \frac{d^3\vec{p}}{(2\pi)^3}\
f_e(E_p,\vec{x})\times\nn \\
&\times & \int \frac{d\Omega_{k'}}{64\pi^2}
\frac{|{\cal M}|^2}{E_p^2(1-\beta\mu)} \frac{I_{\gamma}^0 (t E_{k})}{t}
\;.\label{eq:I_in}
\een
We recall that $t\equiv E_{k'}/E_k$ is the ratio of the diffused to initial
photon energies for the scattered-out process (and conversely for the
scattered-in process), and that the factor of two in front of the right
hand side accounts for the sum over two photon polarizations
(see Eq.~\ref{eq:boltz3}). In the second equality, we have used the expression
of the differential cross section given in Eq.~(\ref{eq:sigma_angle}),
where appears the squared amplitude $|{\cal M}|^2$ of the inverse Compton
process. We further need to average to above expression, as previously done 
for the \emph{out} part, over the solid angle $d\Omega = d\phi d\mu$, and 
the final expression for the scattered-in term is thus:
\ben
\widehat{I}_\gamma^{\rm in}  (E_k)&=& \int\frac{d\Omega_k}{4\pi}\;
I_\gamma^{\rm in}(E_k)\;.
\label{eq:I_in_av}
\een
As one can see, Eqs.~(\ref{eq:I_out_tau}) and (\ref{eq:I_in}) are similar to 
Eqs.~(\ref{eq:app_out}) and ~(\ref{eq:app_in}). While the scattered-out 
processes are strictly equivalent, a slight difference appears in the 
scattered-in, coming from the squared amplitude ($|{\cal{M}}|^2$ versus 
$|\tilde{\cal{M}}|^2$) and the energy transfer ($t$ versus $\tilde{t}$). 
Nevertheless, this difference is completely removed in the relativistic limit, 
since we have in this case $|{\cal{M}}|^2 = |\widetilde{\cal{M}}|^2$ and 
$t = \tilde{t}$. Hence the two methods are indeed equivalent.

\subsection{Redistribution function for the rSZ effect}
\label{subsec:Fmunu}
In Sect.~\ref{subsec:cross_sec}, we will derive the full expression of the 
differential inverse Compton cross section for relativistic electrons and 
soft photons, and also provide some results in the corresponding limit 
$\alpha = E_k/E_p\rightarrow 0$ (see Eq.~\ref{eq:sigma_angle}). Anticipating 
these results, and using Eqs.~(\ref{eq:I_in},\ref{eq:I_in_av}), this implies 
that the angular average of $I_\gamma^{\rm in}(E_k) $ can actually be written 
at leading order in $\alpha$ as:
\ben
\widehat{I}_\gamma^{\rm in}  (E_k)&=& 2 \int dl \int dp f_e(E_p,\vec{x}) \nn\\
&\times &\int d\mu' \int d\mu \ {\cal F}(\beta,\mu,\mu')
\; I_\gamma^0(t E_k)\;,
\label{eq:I_in3}
\een
where we define:
\ben
{\cal F}(\beta,\mu,\mu') &\equiv& \frac{\beta^2 m^2}{(2\pi)^3}
\frac{3\sigma_{\rm T}}{16}
\; \frac{(1-\beta\mu')}{ (1-\beta\mu)^2}  \nn\\
&\times &  \bigg\{ 2 - 2 K(1-\mu\mu') + K^2 \Big\lbrack (1-\mu\mu')^2\nn\\
& & + \frac{1}{2} (1-\mu^2)(1-\mu'^2) \Big\rbrack \bigg\}  \;.
\label{eq:def_fmumup}
\een
$K$ is defined in Eq.~(\ref{eq:sq_amp6}). The function 
${\cal F}(\beta,\mu,\mu')$ is meant to be related, while not strictly, to the 
product $p(\mu) \cdot q(\mu,\mu')$ (where $q(\mu,\mu')$ is in fact the squared 
matrix amplitude) which was displayed in Wright's 
paper~\cite{1979ApJ...232..348W}. However, it is different from Wright's 
expression, notably because this latter is derived in the electron rest frame, 
but they have been shown to be consistent~\cite{2009arXiv0902.2595N}. 

Moreover, we see that in Eq.~(\ref{eq:I_in3}), we cannot factorize out the 
optical depth $\tau$ as defined by Eq.~(\ref{eq:def_tau}): it is mixed with 
the definition of ${\cal F}$, and it cannot be taken out of the integral over 
the angles. Anyway, by re-expressing the photon zenith angles $\mu$ and 
$\mu'$ in the electron rest frame, it is possible to derive a much more 
suitable expression for numerical computation~\cite{2009arXiv0902.2595N}, 
consistent with the results obtained by Wright.

Anyway, the trick one can use here to recover a formulation of the 
\emph{in} process similar to the classical radiative transfer method used by 
Wright is the same as employed for the \emph{out} process. Indeed, we can 
still describe the electron phase space density as $f_e(E_p,\vec{x}) = 
n_e(\vec{x})\tilde{f}_e(E_p)$, where $\tilde{f}_e$ is normalized to unity, and 
utilize the expression of the non-relativistic optical depth $\tau_{\rm nr}$ 
given in Eq.~(\ref{eq:corr_tau_nr}). Armed with this, we can express \emph{in}
contribution \`a la Wright as:
\ben
\label{eq:boltz_in}
\widehat{I}_\gamma^{\rm in}(E_k) &=& 2\;\tau_{\rm nr}\;
\int dp \int d\mu \int d\mu'  \widehat{\cal F} (\beta , \mu , \mu ') \;
I_\gamma^0(t E_k)\;,
\een
where we have defined:
\ben
\widehat{\cal F} (\beta,\mu,\mu') &\equiv &
\frac{{\cal F}(\beta,\mu,\mu')}{\sigma_{\rm T}}\;\tilde{f}_e(E_p) \;.
\label{eq:boltz_fmu}
\een
These are some expressions complementary to those derived by in Wright'. The
function $\widehat{\cal F}$ has to be interpreted as the photon frequency
redistribution function for a given normalized energy distribution of
relativistic electrons such that $\int d^3\vec{p}/(2\pi)^3\tilde{f}_e(E_p)=1$.

To summarize, the Boltzmann approach allows to derive contributions for the
scattered \emph{out} and \emph{in} processes which are similar to the
radiative transfer method when regarding the relativistic SZ effect. These
contributions are recalled here:
\ben
\begin{array}{|ccc|}\hline
\widehat{I}_\gamma^{\rm out}(E_k) &=& 2\;{\cal K}\; \tau_{\rm nr} \;
I_\gamma^0(E_k)\\
\widehat{I}_\gamma^{\rm in}(E_k) &=& 2\;\tau_{\rm nr}\;
\int dp \int d\mu \int d\mu'  \widehat{\cal F} (\beta , \mu , \mu ')
I_\gamma^0(t E_k) \\
\hline \end{array}
\label{eq:boltz_to_rad}
\een
We recall that the factors of two appearing above explicitly account for the
sum over the two polarization states of the CMB photons, so that $I_\gamma^0$
must be defined for one degree of freedom only. Note that for numerical 
computations, the photon frequency redistribution function ${\cal F}$ 
is more suitably expressed when photon zenith angles are re-expressed in the 
electron rest frame, as shown by~\cite{2009arXiv0902.2595N}, to which 
we refer the reader for computation purposes.

\subsection{Experimental signature of the deviation}
\label{subsec:comp_wright}

Finally, since an experiment has a finite angular resolution, the
relevant quantity to use when doing predictions for specific instruments is
the average value of $\Delta I_\gamma(E_k)$ over the solid angle
$\Delta\Omega_{\rm res}$ carried by the experimental resolution, that is:
\ben
\label{eq:res_ave}
\Delta I_\gamma^{\rm obs}(E_k) = \frac{1}{4\pi}\int d\Omega_{\rm res}\
\Delta I_\gamma(E_k)\;,
\een
where the dependence on  $\Omega_{\rm res}$ of the intensity is hidden in the
spatial dependence of the electron population characterized by $f_e(p,\vec{x})$.
More precisely, if the electron density is spherical in the cluster, then it
only depends on the distance $r=|\vec{x}-\vec{x}_{\rm cc}|$ to the cluster
center. This radius $r$ can be related to the angle $\psi_{\rm res}$ scanning
the resolution range through the relation:
\ben
r= \sqrt{l^2 + d^2 - 2 \, d \, l \, \cos \psi_{\rm res} }\;, \nn
\een
where $d$ is the distance from the observer to the cluster center, and $l$ is
the distance as measured along the line of sight.

\section{(Inverse) Compton scattering}
\label{sec:compton}
In this section, we rederive the famous results for the squared matrix
amplitude and the phase space factors associated with the electron-photon
elastic scattering process in the special kinematics of inverse Compton
scattering. The number of references
associated with the Compton scattering is obviously very large, but
one can find some more details in
e.g.~\cite{1971rqt..book.....B,Peskin:1995ev,1982ApJ...254..301B}

\subsection{Kinematics}
\label{subsec:kin}
The kinematics of the process we are interested in is the scattering of an
electron of four-momentum $p$ off a photon of four-momentum $k$, resulting in
final states $p'$ and $k'$ for the electron and the photon, respectively. We
deal with an elastic collision, in which the conservation of four-momentum
holds, so that:
\ben
\label{eq:en_cons}
E_p + E_k = E_{p'}+E_{k'}\;,\\
\vec{p} + \vec{k} = \vec{p}\;' + \vec{k}\;'\;.\nn
\een
We can now derive almost everything we need from these basic equations of
classical collision theory and from the very simple geometry of the
problem~\cite{1950ratr.book.....C}. Note that the coming results will hold in
any Lorentz frame, unless we specify it.

\subsubsection{The scattered photon energy}
\label{subsubsec:en}
The first quantity that we can calculate is the energy $E_{k'}$ of the outgoing
photon after the collision. Starting from Eq.~(\ref{eq:en_cons}), and
substituting $E_{p'} = \sqrt{m^2+p'^2}$, and $\vec{p}\;'=\vec{p}+\vec{k}-
\vec{k}\;'$, we find:
\ben
\label{eq:ekprime}
E_{p'}^2 &=& (E_p + E_k-E_{k'})^2\\
\Leftrightarrow m^2 + (\vec{p}+\vec{k}-\vec{k}\;')^2 &=&
(E_p + E_k-E_{k'})^2\nn\\
\Leftrightarrow E_{k'} &=& t\; E_k \nn\\
{\rm with}\;\; t &\equiv & \frac{(1-\beta\mu)}{(1-\beta\mu')+
  \alpha(1-\Delta)}\;,\nn
\een
where $\beta \equiv p/E_p$ is the electron velocity, where
\ben
\label{eq:def_alpha}
\alpha \equiv \frac{E_k}{E_p}\;,
\een
which will be a quantity very useful to study the different astrophysical
regimes; and where we have defined the cosine of the angle $\theta$ ($\theta'$)
between the incident electron and the initial (outgoing) photon
by $\mu$ ($\mu'$, respectively); and that of the angle $\Theta$ between the
initial and final photon directions by $\Delta$:
\ben
\label{eq:def_angles}
\mu &\equiv& \cos\theta = \frac{\vec{p}\cdot\vec{k}}{|\vec{p}|\cdot|\vec{k} |}\\
\mu' &\equiv& \cos\theta' = 
\frac{\vec{p}\cdot\vec{k}\;'}{|\vec{p}|\cdot|\vec{k}\;'|} \nn\\
\Delta &\equiv& \cos\Theta = 
\frac{\vec{k}\cdot\vec{k}\;'}{|\vec{k}|\cdot|\vec{k}\;'|} \;.\nn
\een
We can already have a look to the limit value of the scattered energy when
the incoming electron energy is much larger than that of the incoming photon
$E_p\gg E_k \Leftrightarrow \alpha\rightarrow 0$, which is typically the case
in any collision between a CMB photon and an electron (indeed, $E_k\ll m$):
\ben
t = \frac{E_{k'}}{E_k} \xrightarrow{\alpha\rightarrow 0}
\frac{(1-\beta\mu)}{(1-\beta\mu')}\;.
\label{eq:ekprime_lim}
\een
We see that, as $\beta\rightarrow 1$, $E_{k'}$ is maximized for frontal
collisions scattering a photon of initial direction $\mu = -1$ to the
opposite direction $\mu' = 1$. In this case, $E_{k'}\rightarrow 4 \gamma^2 E_k$,
obtained after a Taylor expansion of $\beta \simeq 1 - 1/(2\gamma^2)$ up
to the second order in $\gamma\rightarrow\infty$.

\subsubsection{Geometry}
\label{subsubsec:geom}
It will be found convenient to express $\Delta$ as a function of the other
cosines $\mu$ and $\mu'$. Without loss of generality, we can choose the
incident electron direction along the $z$ axis of any frame $Oxyz$.
Therefore, we can define the coordinates of the quadri-momenta as follows:
\ben
\label{eq:geom1}
p &=& E_p(1, 0, 0,\beta) \\
k &=& E_k (1,\sin \theta \cos \phi , \sin \phi  \sin \theta,\cos \theta) \nn\\
k' &=& E_{k'}(1,\sin \theta' \cos \phi',\sin \phi' \sin \theta',\cos \theta')\nn
\;,
\een
where we have used $\beta = |\vec{p}|/E_p$ and the fact that
$|\vec{k}| = E_k$ for photons. The coordinates of $p'$ are readily deduced
from the above expressions by invoking energy-momentum conservation. With
these coordinates, we recover the previous definitions of $\mu$ and $\mu'$
(see Eq.~\ref{eq:def_angles}), and we can now express $\Delta$, the cosine of
the angle between the incoming and outgoing photons, as:
\ben
\Delta &=& \cos\Theta = \mu\mu' + \sqrt{1-\mu^2}\sqrt{1-\mu'^2}\cos\Psi\;,\nn\\
{\rm where}\;\; \Psi &\equiv& \phi - \phi'\;.
\label{eq:geom2}
\een
Armed with these definitions and quantities, we can now survey the full
relativistic phase space as well as the squared amplitude of the process.

\subsection{Phase space}
\label{subsubsec:phase}
The first relevant quantity when dealing with the probability of any
quantum scattering process is the Lorentz invariant phase space $d{\rm LIPS}$,
namely the possible final states to which the process can lead. With the same
conventions as previously, we can write:
\ben
d{\rm LIPS} = \frac{d^3\vec{p}\;'}{(2\pi)^3}
\frac{d^3\vec{k}\;'}{(2\pi)^3 }\;
\frac{(2\pi)^4 \delta^4(p+k-p'-k')}{4E_{p'}E_{k'}}\;.
\label{eq:dlips1}
\een
By performing the integral over $d^3\vec{p}\;'$, we get:
\ben
\int_{\vec{p}\;'}d{\rm LIPS} = \frac{d^3\vec{k}\;'}{4(2\pi)^2} \;
\frac{\delta(E_p+E_k-E_r-E_{k'})}{E_r E_{k'}}\;,
\label{eq:dlips2}
\een
where, accounting for the momentum conservation
$E_{p'}\xrightarrow{\delta^3(P_i-P_f)} E_r$, we define:
\ben
\label{eq:epprime}
E_r &\equiv& \sqrt{m^2 + (\vec{p}+\vec{k}-\vec{k}\;')^2}\\
&=& \sqrt{E_p^2 + E_k^2 +E_{k'}^2 + 2 E_p\beta(E_k\mu-E_{k'}\mu')-2E_kE_{k'}
\Delta}\nn
\een
where the last line is obtained after having expressed $\mu$ as a function of
$t$. Therefore, we see that the argument in the $\delta$ Dirac function of
Eq.(\ref{eq:dlips1}) is a non trivial function of $E_{k'}$,
$f(E_{k'}) = E_p+E_k-E_r-E_{k'}$. By means of the properties of the Dirac
functions, we can write:
\ben
\delta(f(E_{k'})) = \sum_i \frac{\delta(E_k^i - E_{k'})}{| f'(E_k^i)|}\;,\nn
\een
where $E_k^i$ are the zeros of $f$, and $f'$ stands for the derivative with
respect to $E_{k'}$. There is only one zero, $E_{k'}=E_k^0=t\; E_k$, given
by Eq.~(\ref{eq:ekprime}), so that we find:
\ben
\label{eq:dirac}
f'(E_k^0) &=& - \frac{E_p}{E_r^0}
\left( \alpha(t-\Delta) - \beta\mu' + E_r^0/E_p \right)\\
{\rm with}\;\; E_r^0 & \equiv & E_r(E_{k'}=E_k^0)\nn\\
&=& E_p \sqrt{ 1 + 2\alpha\beta(\mu-t\mu') +\alpha^2(1-2 t \Delta+t^2) }\nn\\
&=& E_p | 1+ \alpha (1-t)  |\;.\nn
\een
Therefore, the Dirac function becomes:
\ben
\label{eq:dlips3}
\delta(f(E_{k'})) &=& \frac{E_r^0 \delta(E_k^0 - E_{k'}) }
      {E_p|{\cal B}| }\;,\\
{\rm where} \;\; {\cal B} &\equiv &  \alpha(t-\Delta) - \beta\mu'
+ E_r^0/E_p\nn\\
& = & (1-\beta\mu') + \alpha(1-\Delta)\;.\nn
\een
We can go ahead with the derivation of the phase space factor, which can now be
written as:
\ben
\label{eq:dlipsres}
\int_{E_{k'},\vec{p}\;'}d{\rm LIPS} &=& \int_{E_{k'}}
\frac{d\Omega_{k'}}{4(2\pi)^2} \frac{k'^2dk'}{E_r E_{k'}}
\frac{E_r^0 \delta(E_k^0 - E_{k'}) }
{E_p \; |{\cal B}|}\\
&=& \frac{d\Omega_{k'}}{4(2\pi)^2}\frac{E_k}{E_p}\frac{t}
     {|{\cal B}|}\nn\\
&=& \frac{d\Omega_{k'}}{4(2\pi)^2}\; \frac{\alpha \; t^2}{(1-\beta\mu)}\nn\;.
\een
The second line is obtained after integration over $k'^2dk' = E_{k'}^2 dE_{k'}$.

As previously, it is interesting to consider the limit $\alpha\rightarrow 0$
(see Eq.~\ref{eq:def_alpha}), which is valid in the case of CMB photons
scattered by hot or relativistic electrons ($E_p\gg E_k$). In this limit,
$t\rightarrow (1-\beta\mu)/(1-\beta\mu')$ and we finally get:
\ben
\label{eq:dlipsres_lim}
\int_{E_{k'},\vec{p}\;'}d{\rm LIPS} &\xrightarrow{\alpha\rightarrow 0}&
\frac{d\Omega_{k'}}{4(2\pi)^2}
\; \frac{\alpha (1-\beta\mu)}{(1-\beta\mu')^2}
\een
This result describes the whole phase space function for electron interactions
with CMB photons. It can be expressed in terms of $dt$ or $d\xi$ (where
$\xi\equiv 1/t$), if we make the change of variable $t\leftrightarrow \mu'$ or
equivalently $t\leftrightarrow \mu$. With the definition of $t$ given in
Eq.~(\ref{eq:ekprime}), and in the limit $\alpha\rightarrow 0$ we have:
\ben
\label{eq:dmu_to_dt}
d\mu' &=& \frac{(1-\beta \mu)}{\beta} \; \frac{dt}{t^2}\\
d\mu &=& -\frac{(1-\beta\mu')}{\beta} dt\nn\;.
\een
Thus, in terms of $t$ or $\xi$ , the phase space is merely:
\ben
\label{eq:dmu_dlips}
\int_{E_{k'},\vec{p}\;'} d\mu \; d{\rm LIPS}
&\xrightarrow{\alpha\rightarrow 0}&
\frac{d\phi'}{4(2\pi)^2} \frac{\alpha}{\beta} d\mu \; dt\;,\\
\bigg( &\xrightarrow{\alpha\rightarrow 0}&
- \frac{d\phi'}{4(2\pi)^2} \frac{\alpha \; t}{\beta} d\mu' \; dt \bigg)
\;,\nn\\
{\rm or}\;\; \int_{E_{k'},\vec{p}\;'} d\mu \; d{\rm LIPS}
&\xrightarrow{\alpha\rightarrow 0}&
-\frac{d\phi'}{4(2\pi)^2} \frac{\alpha}{\beta} d\mu \;
\frac{d\xi}{\xi^2}\;,\nn\\
\bigg( &\xrightarrow{\alpha\rightarrow 0}&
 \frac{d\phi'}{4(2\pi)^2} \frac{\alpha}{\beta} d\mu' \; \frac{d\xi}{\xi^3}
\bigg) \;.\nn
\een

\subsection{Squared matrix amplitude}
\label{subsubsec:ampl}
Only two Feynman diagrams have to be considered for the (inverse) Compton
diffusion at leading order. The amplitude of the process
$p+k\rightarrow p'+k'$ is to be found in any textbook introducing quantum
field theory (see e.g.~\cite{Peskin:1995ev}), and can be written as:
\ben
\label{eq:sq_amp1}
{\cal M} &=& -i e^2 \bar{u}_f (p')\;
{\cal A}\; u_i (p) \;,\\
{\rm with}\;\; {\cal A} &=& \myslash{\epsilon}_f^\star
\frac{(\myslash{p}+\myslash{k}+m)}{s-m^2}\myslash{\epsilon}_i +
\myslash{\epsilon}_i \frac{(\myslash{p} - \myslash{k}\;' +m)}{u-m^2}
\myslash{\epsilon}_f^\star\;, \nn
\een
where $i$ and $f$ flag the initial and final states, $p$ and $k$ (prime) are
the four-momenta of the incoming (outgoing) electron and photon, respectively,
$m$ is the electron mass and $u_{i/f}$ is the electron four-spinor, while
$\epsilon_{i/f}$ is the space-like photon four-polarization. The Mandelstam
variables are:
\ben
s &=& (p+k)^2 = m^2 + 2 p.k = m^2 +2 E_p E_k(1-\beta\mu)\nn\\
u &=& (p-k')^2 = m^2 - 2 p.k' = m^2 - 2 E_p E_{k'}(1-\beta\mu')\;.\nn\\
\een
If we define $D_s \equiv s-m^2 = 2 p.k$, $D_u\equiv u - m^2 = -2 p .k'$,
$q_s \equiv p+k$ and $q_u = p - k'$, then the squared value of the amplitude
in Eq.~(\ref{eq:sq_amp1}), once averaged over the incoming electron spin, is
given by:
\ben
\label{eq:sq_amp2}
\frac{1}{4}\sum_{\rm spin,\ pol} |{\cal M}|^2 &=&\frac{1}{2}\sum_{\rm pol} 
\frac{e^4}{2D_s D_u}
Tr\left\{ D_s^2 {\cal T}_1+ D_u^2 {\cal T}_2 + D_sD_u {\cal T}_3 \right\}\\
{\rm with}\;\; {\cal T}_1 &=& (\myslash{p}'+m)\myslash{\epsilon}_i
(\myslash{q}_u+m)\myslash{\epsilon}_f^{\star} 
(\myslash{p}+m)\myslash{\epsilon}_f
(\myslash{q}_u+m)\myslash{\epsilon}_i^{\star} \nn\\
{\cal T}_2 &=& (\myslash{p}'+m)\myslash{\epsilon}_f^{\star} 
(\myslash{q}_s+m)\myslash{\epsilon}_i (\myslash{p}+m) 
\myslash{\epsilon}_i^{\star} 
(\myslash{q}_s+m)\myslash{\epsilon}_f\nn\\
{\cal T}_3 &=& (\myslash{p}'+m)\myslash{\epsilon}_f^{\star} 
(\myslash{q}_s+m)\myslash{\epsilon}_i (\myslash{p}+m) \myslash{\epsilon}_f
(\myslash{q}_u+m)\myslash{\epsilon}_i^{\star} \nn\\
&& + (\epsilon_i\leftrightarrow \epsilon_f \;\&\&\;
q_s \leftrightarrow q_u )\;.\nn
\een
After some Dirac algebra calculation, and after averaging over the photon
polarization states, one can find the very well known result for the squared
Compton amplitude (see e.g.~\cite{Peskin:1995ev}):
\ben
\label{eq:sq_amp3}
|{\cal M}|^2 &=& 2 e^4 \; \bigg\{ \frac{p\cdot k}{p\cdot k'} + 
\frac{p\cdot k'}{p\cdot k} +\\
&& + 2 m^2 \left( \frac{1}{p\cdot k} - \frac{1}{p\cdot k'} \right) +\nn\\
&& + m^4  \left( \frac{1}{p\cdot k} - \frac{1}{p\cdot k'} \right)^2 \bigg\} 
\;.\nn
\een
Using the conventions displayed in Sect.~\ref{subsubsec:geom}, we have
the four-products:
\ben
p\cdot k = m  \ \gamma \ E_{k} \ (1 - \beta \mu)
\label{eq:pk}
\een
and
\ben
p\cdot k' = m \ \gamma \ E_{k'} \ (1 - \beta \mu')
\label{eq:pkp}
\een
so Eq.~(\ref{eq:sq_amp3}) finally reduces to:
\ben
|{\cal M}|^2 &=& 2 e^4 \;
\big\{  1+
\left( 1-\frac{(1-\Delta)}{\gamma^2(1-\beta\mu)(1-\beta\mu')}\right)^2+\nn\\
&& + \frac{\epsilon^2(1-\Delta)^2 }{\gamma^2 (1-\beta\mu')
\left[ (1-\beta\mu') + \alpha(1-\Delta)\right]} \big\} \nn\\
&=& 2 e^4 \; \left\{ 2 + \frac{g}{\tilde{\gamma}^2 } \;
\left(\frac{g}{\tilde{\gamma}^2 }- 2 \right)+ \frac{\alpha^2 g^2}{1+\alpha g}
\right\}\;,
\label{eq:sq_amp4}
\een
where we define:
\ben
g &\equiv& (1-\Delta)/(1-\beta\mu')\nn\\
{\rm and}\;\; \tilde{\gamma}^2 &\equiv & \gamma^2 (1-\beta\mu)\;.
\een
$\gamma \equiv E_p/m$ is the usual Lorentz factor for the incoming electron,
and $\epsilon\equiv E_k/m = \alpha\gamma$ is the incoming photon energy in
units of the electron mass. The second equality given in Eq.~(\ref{eq:sq_amp4})
is very convenient because the result of the limit
$E_p\gg E_k \Leftrightarrow \alpha\rightarrow 0$ appears explicitly. Within
this limit, the squared amplitude is merely:
\ben
|{\cal M}|^2 \xrightarrow{\alpha\rightarrow 0} 2 e^4 \;
\left\{ 2 + \frac{g}{\tilde{\gamma}^2 } \;
\left(\frac{g}{\tilde{\gamma}^2 } - 2 \right) \right\}\;.
\label{eq:sq_amp5}
\een
Should we implement the non-relativistic limit $\beta\rightarrow 0$, we
would immediately recover the well known angular dependence of the Thomson
scattering, that is $|{\cal M}|^2\propto 1+\Delta^2$. Anyway, the above
expression depends on the azimuthal angle $\Psi$ through $\Delta$ (see
Eq.~\ref{eq:geom2}), and it is interesting to perform the azimuthal integral
for deeper insights. It is clear that only the even powers of $\cos\Psi$ will
contribute to this integral. We find:
\ben
\label{eq:sq_amp6}
\langle |{\cal M}|^2 \rangle_\Psi &\xrightarrow{\alpha\rightarrow 0}&
2 e^4 \; \bigg\{ 2 - 2 K(1-\mu\mu') + \\
& & + K^2 \left\lbrack [(1-\mu\mu')^2 +
\frac{1}{2} (1-\mu^2)(1-\mu'^2) \right\rbrack \bigg\} \nn\\
{\rm with} \;\; K &\equiv & \left[ \tilde{\gamma}^2
(1-\beta\mu')\right]^{-1} \;.\nn
\een
From this expression, it is easy to calculate the limit $\beta\rightarrow 0$,
that is the expression corresponding to the scattering of a photon
of energy $E_k\ll m$ off an electron at rest. Indeed, within this limit,
$K\rightarrow 1$, and the squared amplitude becomes:
\ben
\label{eq:sq_amp7}
\langle |{\cal M}|^2 \rangle_\Psi
&\xrightarrow{\beta\rightarrow 0}&
2 e^4 \; \left\{1 + \frac{1}{2}(1-\mu^2)(1-\mu'^2) + \mu^2\mu'^2\right\}\;.
\een
This is the famous non-relativistic limit derived by Chandrasekhar in his
textbook~\cite{1950ratr.book.....C} and used in most of treatments of SZ
computation, and in particular in~\cite{1979ApJ...232..348W} (by Lorentz
boosting the photons up and down with respect to the electron rest frame).
This expression can not be used such as it is written above in relativistic
SZ calculations, but it turns out that a form very similar can be derived 
when expressing the photon zenith angles $\mu$ and $\mu'$ in the 
electron rest frame, that is $\mu = (\beta-\mu_0)/(1-\beta\mu_0)$ and 
$\mu' = (\beta-\mu_0')/(1-\beta\mu_0')$, as nicely detailed 
in~\cite{2009arXiv0902.2595N}.

\subsection{Differential (inverse) Compton cross section}
\label{subsec:cross_sec}

The (inverse) Compton differential cross section is defined as usual for two
body elastic scattering $(p,k\rightarrow p',k')$:
\ben
d^2\sigma = \frac{(2\pi)^4}{4 p\cdot k} \left\{ \prod_{X=p',k'}
\frac{d^3\vec{X}}{(2\pi)^32 E_X} \right\} \delta^4(p+k-p'-k')|{\cal M}|^2 \;,
\label{eq:def_sigma}
\een
where $|{\cal M}|^2$ already includes the sum and average over polarizations,
and where the 4-product $p\cdot k = E_p E_k \beta_{\rm rel}$ in terms of the 
energies of
the incoming particles and their associated relative velocity. After
integration over $\vec{p}\;'$ and $E_{k'}$, by virtue of the energy-momentum
conservation, the angular differential cross section reads:
\ben
\frac{d\sigma}{d \Omega_{k'}} = \frac{1}{4E_k E_p\; \beta_{\rm rel}}
\int_{\vec{p}\;',E_{k'}} d{\rm LIPS} \; |{\cal M}|^2\;.
\label{eq:def_sigma2}
\een
After substituting $\beta_{\rm rel}= (1- \beta \mu)$ (see Eq.~\ref{eq:pk}):
\ben
\label{eq:sigma_angle}
\frac{d\sigma}{d\Omega_{k'}} &=&
\frac{1}{64\pi^2 }\frac{t^2}{E_p^2 (1-\beta\mu)^2} |{\cal M}|^2 \\
&=& \frac{3\sigma_{\rm T}}{16\pi}\frac{t^2}{\gamma^2 (1-\beta\mu)^2}
\frac{|{\cal M}|^2}{2 e^4}\nn\\
& \xrightarrow{\alpha \rightarrow 0} &
\frac{3 \sigma_{\rm T}}{16 \pi \gamma^2(1-\beta\mu')^2}
\bigg\{ 2 - 2 K(1-\mu\mu') + \nn\\
& & + K^2 \left\lbrack [(1-\mu\mu')^2 +
\frac{1}{2} (1-\mu^2)(1-\mu'^2) \right\rbrack \bigg\}  \nn\;,
\een
where we have used the standard definition of the Thomson cross section
$\sigma_{\rm T} = 8\pi r_0^2/3$, with the classical electron radius
$r_0\equiv e^2/(4\pi m) = \alpha_{\rm fs}/m$. The first two lines give the
exact and general expression of the (inverse) Compton scattering differential
cross section. The full relativistic expressions of $t$ and $|{\cal M}|^2$
have been derived in Eqs.~(\ref{eq:ekprime}) and (\ref{eq:sq_amp4}),
respectively. The third line shows the relativistic limit at leading order,
when $\alpha = E_k/E_p\rightarrow 0$, including the average over the photon
azimuthal angle, according to Eq.~(\ref{eq:sq_amp6}) where $K$ is also
defined. If we further integrate\footnote{We temporarily slightly change our
convention which defines the cosine $\Delta$ with respect to $\mu$ and
$\mu'$ --- see Eq.~(\ref{eq:def_angles}). We now set the main axis of the
problem along $\vec{k}\;'$ instead of $\vec{p}$, so that it is now $\mu$ which
depends on $\Delta$ and $\mu'$ and so that we can perform the integral over
$d\mu'$ without accounting for $\Delta$. Such a trick is aimed at comparing
our results with the standard calculation of the integrated Compton cross
section.} the full differential cross section over $d\Omega_{k'}$, we find
(see e.g.~\cite{1971rqt..book.....B,1982ApJ...254..301B}):
\ben
\label{eq:sigma_int}
\sigma(E_p,E_k) &=& \frac{3 \sigma_{\rm T}}{4}\bigg\{
\frac{(\chi^2-2\chi-2)}{2\chi^3}\ln(1+2\chi)\\
&& + \frac{(1+2\chi)^2 + (1+\chi)^2 (1+2\chi) - \chi^3 }{\chi^2(1+2\chi)^2}
\bigg\}\nn\\
&\xrightarrow{\alpha\gamma^2\rightarrow 0}& \sigma_{\rm T}\;,\nn
\een
where $\chi\equiv \alpha \gamma^2(1-\beta\mu)$. We have featured the limit
when $\alpha\gamma^2 = \gamma E_k/m \rightarrow 0$. It makes sense to
recover the Thomson cross section, since if we transpose the process in
the electron rest frame, the photon would have an energy of $\gamma E_k$,
still much lower than $m$ as long as $\gamma\lesssim 10^8$, which corresponds
to the classical Thomson regime. While not describing a Thomson scattering
at all, the corresponding \emph{integrated} cross section is recovered as
long as $\gamma E_k\ll m$: this is a mere consequence of special relativity,
and very well know for decades in
astrophysics~\cite{1965PhRv..137.1306J,1968PhRv..167.1159J,1970RvMP...42..237B}.

Another useful quantity to derive is the integral over the incoming and
outgoing photons angles of the Taylor expansion of
$\beta_{\rm rel} d\sigma/d\Omega_{k'}$ in the limit $\alpha\rightarrow 0$. Up
to the second order in $\alpha$, we get:
\ben
\int d\mu \int d\mu' \int d\phi ' \beta_{\rm rel}
\frac{d\sigma}{d\Omega_{k'}} \xrightarrow{\alpha\rightarrow 0}
2\sigma_{\rm T}  \bigg\{ && 1 + \frac{2\alpha}{3}
( 1 - 4 \gamma^2 ) -\nn\\
&& \frac{26\alpha^2}{5} (\gamma^2 - 2 \gamma^4) \bigg\}\;,\nn\\
\label{eq:lim_sigma_int}
\een
where we have expressed the cosine $\Delta$ as a function of $\mu$, $\mu'$
and $\phi'$ according to Eq.~(\ref{eq:def_angles}) before performing the
integral.

\section{Conclusion}
\label{sec:concl}

We have shown that the approach that Wright developed
in Ref.~\cite{1979ApJ...232..348W} was equivalent to using a Boltzmann-like
equation in the single scattering approximation, already very well
described in the literature (e.g.~\cite{1998ApJ...499....1C,
1998ApJ...502....7I,1997astro.ph..9065S,2001ApJ...554...74D}). The 
same result has been recovered in a recent subsequent work by Nozawa 
and Kohyama~\cite{2009arXiv0902.2595N}. Here, we have established the 
relativistic expressions valid in the CMB frame to employ within Wright's 
formalism, summarized in Eqs.~(\ref{eq:boltz_to_rad}). They differ from 
Wright's expressions which are instead derived in the electron rest 
frame~\cite{2009arXiv0902.2595N}, and which are probably more suited for 
numerical computations. Finally, we have recovered that the SZ shift at energy 
$E_k$ due to a relativistic population of electrons is given by an integral 
over a function which is still proportional to 
$I_{\gamma}^0 (E_k)-I_{\gamma}^0(t\; E_k)/t^3$, where ${I_{\gamma}}^0$ is the
intensity of the pure black-body CMB spectrum and $t = E_{k'}/E_k$ is the
ratio of scattered-to-focused frequencies.

\section*{Acknowledgements}
We would like to thank M. Giovaninni for illuminating discussions on
Chandrasekhar's work and P. Salati for interesting remarks. We are also
indebted to E. Wright, N. Itoh and S. Nozawa for their valuable comments 
after the paper was put on the arXiv, pointing out, in particular, 
a mistake that we did in the former version of this paper on the 
interpretation of E. Wright's early results. CB is grateful to 
CERN IT for providing computer support and CERN TH division for hospitality.

\appendix
\section{Recall of the radiative transfer approach}
\label{app:wright_formalism}
In this appendix, we would like to demonstrate that the radiative transfer
method used by Wright in~\cite{1979ApJ...232..348W} can be deduced from the
Boltzmann approach presented in this paper, but by starting from the
radiative approach instead. This will provide more insights on the correct
frequency redistribution function to be used for the CMB photons. 

We are looking for a redistribution function that expresses the probability
$P(t)$ for a CMB photon to have its energy $E_k$ shifted to $E_{k'}$ by a
factor of $t\equiv E_{k'}/E_{k}$ --- we could equivalently deal with
$P(s)$ with $s\equiv \ln(t)$ ---  in order to be able to write
(see e.g.~\cite{1999PhR...310...97B}) the modified intensity as:
\ben
I_\gamma(E_k) = \int
dt\; P(t)\times I^0_\gamma(t E_k)\;.
\label{eq:rad_t_meth}
\een
The integral is performed from the minimal value of the scattered
energy $t_{\rm min} = 1/(4\gamma_{\rm max}^2-1)$ up to its maximal value
$t_{\rm max} = (4\gamma_{\rm max}^2-1)$, as allowed by the kinematics.

According to the radiative transfer method, such a probability depends
on the electron optical depth $\tau$ and on the frequency redistribution
function $P_1(t)$ for a single scattering. In the limit of single scattering,
it is merely given by:
\ben
P(t) &=& e^{-\tau}\left\{ \delta(t-1) + \tau P_1(t) \right\}\nn\\
&\xrightarrow{\tau \ll 1}& (1-\tau) \delta(t-1) + \tau P_1(t) +O(\tau^2)\;,
\label{eq:def_pt}
\een
where, in the second line, we have taken the limit of very small optical depth,
which is the case in the problem of interest here, but also more generally
for electrons in clusters.

If we report the previous equation in Eq.~(\ref{eq:rad_t_meth}), we find:
\ben
\Delta I_\gamma (E_k) = \tau \int dt\; \left( P_1(t) - \delta(t-1) \right)
\times I_\gamma^0 (t E_k)\;.
\label{eq:deltaI_rad_t_meth}
\een
Two terms appear in the sum in the integral, which are readily interpreted
as the positive contribution of CMB photons scattered \emph{in} from energy
$E_{k'} = tE_k$ to energy $E_k$ for the former, and as the negative
contribution of photons which have been shifted \emph{out} from energy $E_k$ to
other energies. More explicitly, those contributions are given by:
\ben
I_\gamma^{\rm out}(E_k) \equiv \tau \times I_\gamma^0 (E_k)\;,
\label{eq:Iout_rad_t_meth}
\een
and:
\ben
I_\gamma^{\rm in}(E_k) \equiv \tau \int dt\; P_1(t) \times I_\gamma^0 (t E_k)\;.
\label{eq:Iin_rad_t_meth}
\een
If we scrutinize Eqs.~(\ref{eq:boltz_out}) and~(\ref{eq:boltz_in}), we can find
some convincing similarities, while not straightforwardly. With the
definition of $\tau$ given in Eq.~(\ref{eq:def_tau}), the former is
perfectly consistent with Eq.~(\ref{eq:Iout_rad_t_meth}). Nevertheless,
it is a bit more tricky for the latter. Indeed, given our definition,
because $\tau$ already includes the integral over $t$, it would have to be
considered as an operator acting on $I_\gamma$. We have actually determined
the full relativistic formula, summarized in Eqs.~(\ref{eq:boltz_to_rad}).

\section{Scattered \emph{in} and \emph{out} photons treated separately}
\label{app:in_out}

\subsection{Definitions}

Let us start with some basic definitions. The energy density of the black-body
spectrum, of average temperature $T_0$, before traveling through the cluster
is given by:
\ben
\rho_\gamma^0 =  \int d^3 \vec{k} \ E_k \ f_{\gamma}^0(E_k),\nn
\een
with $d^3 k  = E_k ^2 \ dE_k  \ d \Omega_k$ and
\ben
f_{\gamma}^0(E_k) = \frac{1}{e^{E_{k}/T_0} -1}\;.\nn
\een
From this equations, we deduce that the associated intensity is:
\ben
\label{eq:Ik1}
I^0_\gamma(E_k) &=& E_{k}^3 \ f_{\gamma}^0(E_k)  \\
&=& \ \frac{E_k ^3}{e^{E_k/T_0 }-1}. \nn
\een

\subsection{Intensity associated with the relativistic effect}
\label{subsec:appendix}
When interacting with the relativistic electrons, some CMB photons initially
at energy $E_k$ can be scattered out at an energy $E_{k'}$. This effect
generates a deficit in the number of CMB photons that is expected at an energy
$E_k$. However, other CMB photons with an energy $E_{k'}\neq E_k$ will be
scattered in the cluster  to a final energy $E_k$. The photons associated
with the former process (responsible for the deficit) will be referred to as
"scattered out" photons (and will be denoted hereafter "s-out") while those
associated with the second process will be referred to as "scattered in
photons" (and will be denoted "s-in"). Let us now examine these two processes
separately.

\subsubsection{Intensity associated with the photons scattered out}

After traveling through the cluster and interacting with the population of
relativistic electrons, a CMB photon (with an initial energy $E_k$) can be
shifted to a new energy $E_{k'}\neq E_k$, according to the reaction
$k + p \rightarrow k' + p'$. The CMB energy spectrum is modified
accordingly. For an observer who aims at measuring the resulting intensity of
the CMB spectrum, such interactions with a population of relativistic
electrons translates into a deficit in the number of photons at an energy
$E_k$ with respect to the black-body distribution.

If we disregard the thermal and kinetic SZ effect for a moment, the number of
photons which remain at this energy $E_k$ after traveling through the cluster
is the number of photons which had this energy before entering the cluster
minus the number of photons $n_{\gamma}^{\rm s-out}$ which have changed of
energy after traveling through the cluster.

This number density $n_{\gamma}^{\rm s-out}$ of scattered photons is basically
the convolution over the 3-momentum $d^3\vec{k}$ between the initial black-body
occupation number $f_\gamma^0$ and the interaction rate with the electrons
(that is $\Gamma_{\gamma} = n_e \sigma \beta_{\rm rel}$ if
$\sigma \beta_{\rm rel}$ is independent of the ingoing electron energy $E_p$).
For relativistic electrons, the dependence of the cross section on the energy
$E_p$ is not trivial and the interaction rate is rather given by:
\ben
\Gamma_\gamma(E_k) =   \int \frac{d^3\vec{p}}{(2 \pi)^3} \; f_{e}(E_p)
\; [\sigma \beta_{\rm rel}]_{p k\rightarrow p' k'} \nn
\een
where
$$[\sigma \beta_{\rm rel}]_{p k\rightarrow p' k'} = \frac{1}{4 E_p \ E_k }
\int_{p',k'} d{\rm LIPS} \; |{\cal M}|^2 $$
with $\beta_{\rm rel}$ the relative velocity and where the Lorentz invariant
phase space $d{\rm LIPS} = d{\rm LIPS}_{p',k'}$ is defined in
Eq.~(\ref{eq:dlips1}), and further developed in Eq.~(\ref{eq:dlipsres}).
$|{\cal M}|^2 = |{\cal M}|^2_{pk\rightarrow p'k'}$ is given in
Eq.~(\ref{eq:sq_amp4}). To derive the full distribution function of the
photons which have been scattered out from energy $E_k$ to energy $E_{k'}$, we
have to multiply (or weight) the black-body distribution by the interaction
rate integrated over the time:
\ben
f_{\gamma}^{\rm{s-out}}(E_{k}) &=&f_{\gamma}^0(E_k)\int dt\;\Gamma_\gamma(E_k)\nn\\
&=& f_{\gamma}^0(E_k) \int dt\int \frac{d^3 p}{(2 \pi)^3} \; f_{e}(E_p)
\;  [\sigma \beta_{\rm rel}]_{k p\rightarrow k' p'} ,\nn
\een
which provides the full number density of photons scattered-out from any energy
$E_k$ to any $E_{k'}$, that is the number density of photons having interacted:
\ben
n_\gamma^{\rm s-out} =   \int \frac{d^3 k}{(2 \pi)^3}  \;
f_{\gamma}^{\rm{s-out}}(E_{k})\;.\nn
\een
Now, we are interested in the specific density removed from a given
energy $E_k$ and shifted to any energy $E_{k'}$, so we have to come back to
the distribution function. Hence, the intensity associated with the
scattered-out photons reads:
\ben
I_{\gamma}^{\rm{s-out}} (E_{k}) \equiv E_{k}^3\ f_{\gamma}^{\rm{s-out}}(E_{k})\nn\;,
\een
where $f_{\gamma}^{\rm{s-out}}(E_{k})$ represents the distribution function
associated with the photons which have experienced elastic scatterings with
electrons in the cluster. Since an observer will detect a modification of the
photon distribution along the line-of-sight, the deficit in the CMB intensity
at an energy $E_k$ that could be observed is merely obtained by using the
geodesic relation $dt = dl/c$ ($= dl$ in natural units):
\ben
\frac{\widehat{I}^{\rm s-out}(E_k)}{E_k^3} &=& \int dl
\int \frac{d^3\vec{p}} {(2 \pi)^3} f_e(E_p) \int_{p',k'} d{\rm LIPS}
\frac{ |{\cal M}|^2 }{4 E_p E_k}  f_{\gamma}^0(E_k) \nn\\
&=&  \int dl
\int \frac{d^3\vec{p}} {(2 \pi)^3} f_e(E_p) \int \frac{d\Omega_{k'}}{64\pi^2}
\frac{t^2|{\cal M}|^2}{(1-\beta\mu)} \frac{f_{\gamma}^0(E_k)}{E_p^2} \nn\;,
\een
where the Lorentz invariant phase space has been made explicit. We have finally:
\ben
\boxed{\widehat{I}^{\rm s-out}(E_k) =  \int dl
\int \frac{d^3\vec{p}} {(2 \pi)^3} f_e(E_p) \int \frac{d\Omega_{k'}}{32\pi^2}
\frac{t^2|{\cal M}|^2}{(1-\beta\mu)} \frac{I_{\gamma}^0(E_k)}{E_p^2} }\nn\\
\label{eq:app_out}
\een
where we used the fact that $I_{\gamma}^0(E_k) =  2 E_k^3 \;
f_{\gamma}^0(E_k)$, making explicit the factor of two coming from the sum over
the photon polarization states, and where the squared
amplitude $|{\cal M}|^2 = |{\cal M}|^2_{p,k\rightarrow p',k'}$. We exactly
recover the scattered-out term as derived in the Boltzmann-like
formalism (see Eqs.~\ref{eq:I_out_tau} and~\ref{eq:def_tau}). Of course,
one will have to average over the solid angle of the incoming photons
$d\Omega_k = d\phi d\mu$.

\subsubsection{Intensity associated with the photons scattered in}
Let us now estimate the number of photons with an arbitrary energy $E_{k'}$
which have actually been scattered to an energy $E_{k}$ by a population of
electrons of energy $E_p$. A first approach consists in doing as in the
previous subsection, by defining the interaction rate of scattering shifting
photons of energy $E_{k'}$ to any energy $E_k\neq E_{k'}$;
\ben
\Gamma_\gamma(E_{k'}) =   \int
\frac{d^3 \vec{p}}{(2\pi)^3} \; f_{e}(E_p)
\; \lbrack \sigma \beta_{\rm rel} \rbrack_{p k'\rightarrow p'k} .\nn
\een
The number density of scattered-in photons from any energy $E_{k'}$ to
any energy $E_k$ is therefore given, in terms of the associated distribution
function, by:
\ben
n_\gamma^{\rm s-in} &=& \int\frac{d^3\vec{k}'}{(2\pi)^3}
f_\gamma^{\rm s}(E_{k'})\nn\\
f_\gamma^{\rm s}(E_{k'}) &\equiv& f_\gamma^0(E_{k'})
\int dt \Gamma_\gamma(E_{k'})\;.\nn
\een
Nevertheless, this is the number density of photons scattered to \emph{any}
energy $E_k\neq E_{k'}$, that is the total number density of photons having
interacted, while we are only interested in the scattered
photons from \emph{any} energy $E_{k'}$ to a \emph{given} energy $E_k$.
Therefore, we have to modify a bit the above picture, and use differential
quantities. The differential interaction rate of shifting photons from energy
$E_{k'}$ to energy $E_k$ reads:
\ben
\frac{d\widetilde{\Gamma}_\gamma(E_{k'},E_k)}{d^3\vec{k}} =   \int
\frac{d^3 \vec{p}}{(2\pi)^3} \; f_{e}(E_p)
\; \beta_{\rm rel} \frac{d\sigma_{p k'\rightarrow p'k}}{d^3\vec{k}} .\nn
\een
To derive the distribution function associated with all photons scattered
from energy $E_{k'}$ to energy $E_k$, we need to convolve the above
differential rate with the initial black-body function and integrate over time
and over all energies $E_{k'}$:
\ben
f_{\gamma}^{\rm s-in}(E_k) &=&  \int\frac{d^3\vec{k}'}{(2\pi)^3} f_\gamma^0(E_{k'})
\int dt \frac{d\widetilde{\Gamma}_\gamma(E_{k'},E_k)}{d^3\vec{k}}\nn\\
&=& \int \frac{d^3 \vec{k}'}{(2 \pi)^3}
f_{\gamma}^0(E_{k'}) \int \frac{d^3 p}{(2 \pi)^3}
f_{e}(E_p) \; \beta_{\rm rel} \frac{d\sigma_{p k'\rightarrow p'k}}{d^3\vec{k}}
\;.  \nn
\een
We have:
$$ \beta_{\rm rel} \frac{d\sigma_{p k'\rightarrow p'k}}{d^3\vec{k}} =
\frac{(2\pi)^4 }{4 E_{k'} E_p} \int\frac{d^3\vec{p}'}{(2\pi)^3} \;
\delta^4(p+k'-p'-k)\frac{|\widetilde{\cal M}|^2}{4 E_k E_{p'}} $$
where $|\widetilde{\cal M}|^2 = |{\cal M}|^2_{p k'\rightarrow p'k}$.
We see that the phase space is more subtle than in the scattered-out term.
We can actually perform the integral over $d^3\vec{k}'$ and $d^3\vec{p}'$,
by means of the conservation of energy-momentum. This will differ a bit
from the scattered-out case, but we will apply the same treatment as in
Sect.~\ref{subsubsec:phase}, to which we refer the reader for more details.
We define:
\ben
d\widetilde{\rm LIPS}  &=& \frac{(2\pi)^4 }{4E_k E_{p'}}
\frac{d^3\vec{p}'}{(2\pi)^3}\frac{d^3\vec{k}'}{(2\pi)^3}
\delta^4(p+k'-p'-k)\;.\nn
\een
This is not the common Lorentz invariant phase space for the process
$p k'\rightarrow p' k$, which should instead feature integrals over $p'$ and
$k$, but this is a useful quantity in the present calculation. If we perform
the integral over $d^3\vec{p}'$, we get:
\ben
\int_{\vec{p}'}d\widetilde{\rm LIPS} &=&
\frac{d^3\vec{k}'}{16\pi^2 E_k \tilde{E}_r}
\delta(E_p+E_{k'} - \tilde{E}_r - E_k)\nn
\een
with:
\ben
\tilde{E}_r &\equiv& E_p \sqrt{1+\alpha'^2+\alpha^2 +
  2\beta(\alpha'\mu'-\alpha\mu)-2\alpha\alpha'\Delta}\;,\nn
\een
where $\alpha\equiv E_{k}/E_p$, $\alpha'\equiv E_{k'}/E_p$, and the other
angular variables are already defined in Eq.~(\ref{eq:def_angles}). We
can further transpose the Dirac $\delta$ function according to:
\ben
\delta(E_p+E_{k'} - \tilde{E}_r - E_k) &=&
\frac{\tilde{E}_r^0\delta(E_{k'} - E_k^0)}{E_p|\widetilde{\cal B}| }\;,\nn
\een
with:
\ben
\label{eq:eqtt}
E_k^0 &\equiv& \tilde{t} E_{k}\nn \\
\tilde{t} &\equiv& \frac{(1-\beta\mu)}{(1-\beta\mu')+\alpha(\Delta-1)}\nn\\
\widetilde{\cal B} &\equiv & (1-\beta\mu') +\alpha (\Delta-1) \nn\\
\tilde{E}_r^0 & \equiv & \tilde{E}_r(E_k^0) = E_p |1-\alpha(1-\tilde{t})| \;.\nn
\een
Hence, performing the integral over $d^3\vec{k}'$, we can rewrite the
phase space as:
\ben
\int_{\vec{p}',E_{k'}}d\widetilde{\rm LIPS} &=&
\frac{d\Omega_{k'}}{16\pi^2}\; \frac{\alpha \tilde{t}^3}{(1-\beta\mu)}\;,\nn
\een
such that the scattered-in distribution function can be written as:
\ben
f_{\gamma}^{\rm s-in}(E_k) &=& \int dt \int \frac{d^3 p}{(2 \pi)^3}
f_e(E_p) \int \frac{d\Omega_{k'}}{64\pi^2} \frac{|\widetilde{\cal M}|^2}{E_p^2}
\frac{\tilde{t}^2 f_\gamma^0(\tilde{t} E_k)}{(1-\beta\mu)}  \nn
\een
We can now derive the intensity of the scattered-in photons at energy $E_k$:
\ben
\widehat{I}^{\rm s-in} (E_{k}\leftarrow E_{k'}) = E_{k}^3
f_\gamma^{\rm s-in}(E_k)\;.\nn
\een
If we transform the time integral into a line of sight
integral, we finally get:
\ben
\boxed{\widehat{I}^{\rm{s-in}} (E_{k}) =
\int dl \int \frac{d^3 p}{(2 \pi)^3}
f_e(E_p) \int \frac{d\Omega_{k'}}{32\pi^2} \frac{|\widetilde{\cal M}|^2}{E_p^2}
\frac{I_\gamma^0(\tilde{t} E_k)}{\tilde{t}(1-\beta\mu)} }\nn\\
\label{eq:app_in}
\een
where we have used $I_\gamma^0(\tilde{t} E_k) =2 (\tilde{t} E_k)^3 f
_\gamma^0(\tilde{t} E_k) $ --- making explicit the factor of two coming from
the sum over the photon polarization states --- and where the squared amplitude
$|\widetilde{\cal M}|^2 = |{\cal M}|^2_{p k' \rightarrow p' k}$. Of course, we
will further need to perform the averaging over the solid angle
$d\Omega_k=d\phi d\mu$ associated with the detected photon. Nevertheless, it
is already interesting to note that we recover an expression which is similar
to Eq.~(\ref{eq:I_in}), which is the result for the \emph{in} process obtained
in the Boltzmann-like formalism, but with slight differences. Indeed,
in the Boltzmann approach, the squared amplitude is
$|{\cal M}|^2_{p k \rightarrow p' k'}$, and the expression of $t = E_{k'}/E_k$
slightly differs from that of $\tilde{t}$. As regards the energy ratios, the
difference comes from the denominators: there is a factor of $\alpha(1-\Delta)$
in this of the former, and of $-\alpha(1-\Delta)$ in that the latter. The 
physical interpretation of this slight difference is rather simple. Indeed, 
$\alpha = E_k/E_p$ in both cases, but though $E_k$ is the energy of the 
incoming photon in the former case, it is that of the outgoing photon in 
the latter case. Nevertheless, since $\alpha\rightarrow 0$ for CMB photons as 
scattered by relativistic electrons, this has no effect. Concerning the 
squared amplitudes, we can also verify that $|{\cal M}|^2$ and 
$|\widetilde{\cal M}|^2$ are equivalent within this limit. There is
therefore no difference between the scattered-in contribution as computed in
the approach developed in this appendix and that computed within the
Boltzmann-like formalism in the limit $\alpha\rightarrow 0$, which fully
applies for relativistic electrons.

\bibliography{lavalle_bib}

\end{document}